\documentclass[prl,nofootinbib,floats,aps,superscriptaddress,twocolumn]{revtex4}

\usepackage{graphicx,amssymb,amsmath}
\usepackage{hyperref}

\newcommand{\beq}{\begin{equation}}
\newcommand{\eeq}{\end{equation}}

\newcommand{\dy}{\Delta y}

\newcommand{\ad}{A^{\rm{\ttbar}}}
\def\bea{\begin{eqnarray}}
\def\eea{\end{eqnarray}}

\newcommand{\pt}{p_T} 
\newcommand{\ttbar}{t\bar{t}}

\newcommand{\mbl}{m_{b\mu}}
\newcommand{\mxl}{m_{xl}}
\newcommand{\Aet}{A^{t\bar t}_{\eta}}

\def\mysection#1{{{\bf #1}.~}}

\usepackage{color}
\definecolor{nicered}{rgb}{0.7,0.1,0.1}

\begin{document}

\title{Top LHCb Physics}

\author{Alexander  L. Kagan}
 \affiliation{Department of Physics, University of Cincinnati, Cincinnati, Ohio 45221, USA}
 \affiliation{Department of Particle Physics and Astrophysics, Weizmann Institute of Science, Rehovot 76100, Israel}

\author{Jernej F. Kamenik} 
\affiliation{J. Stefan Institute, Jamova 39, P. O. Box 3000, 1001 Ljubljana, Slovenia; 
Department of Physics, University of Ljubljana, Jadranska 19, 1000 Ljubljana, Slovenia}

\author{Gilad Perez} 
\affiliation{Department of Particle Physics and Astrophysics, Weizmann Institute of Science, Rehovot 76100, Israel}

\author{Sheldon Stone} 
\affiliation{Department of Physics, Syracuse University Syracuse, N. Y., U. S. A., 13244}

\begin{abstract}
We suggest that top physics can be studied at the LHCb
experiment, and that top production could be observed. Since
LHCb covers a large pseudorapidity region in the forward
direction, it has unique abilities to probe new physics in
the top sector. Furthermore, we demonstrate that LHCb may be
able to measure a $t\bar t$ production rate asymmetry, and
thus indirectly probe Êan anomalous forward backward $t\bar
t$ asymmetry in the forward region; a possibility suggested
by the enhanced forward-backward asymmetry reported by the
CDF experiment.\end{abstract}

\maketitle
\mysection{Introduction}
In the Standard Model (SM) the top quark induces the most severe hierarchy problem. Furthermore, in most natural models it is linked to electroweak symmetry breaking. Consequently, there is strong motivation to search for new physics (NP) effects associated with top physics.
Recently, the CDF collaboration has reported evidence for a large forward-backward asymmetry in top quark pair production~\cite{CDFAFB}.
Interestingly, there is also an indication that the asymmetry increases significantly with rapidity difference,
\beq
   \ad_{\dy>1}   =   \frac{N(\dy > 1) - N(\dy< -1)}{N(\dy > 1) + N(\dy < -1)}  =0.611\pm 0.256  \,,    
\eeq
where N is the number of events with a given rapidity difference $\dy $ between the top and the anti-top.
This result motivates extensions of the SM that enhance top quark production in the very forward region. 
In fact, it is well known that this is a feature of a wide
class of new physics models, i.e, those in which top
production proceeds via t-channel exchange of a new low mass
particle, due to forward peaking in the differential cross section. Regardless of any specific theoretical scenario it is very important to have an experimental probe of this forward region.
Below we argue that the LHCb experiment, by virtue of its high pseudo-rapidity detector capabilities, may provide a unique opportunity for such a study. 
The LHCb detector is far from being hermetic. Thus, substantial event information, {\it e.g.}, missing energy,  is not available.   
Nevertheless we propose that the LHCb detector can study SM top pair production. In addition, it may be sensitive to new physics dynamics where the rate for top pair or single top production is enhanced in the forward direction.
We further demonstrate that this would allow for a probe of the top quark forward-backward asymmetry in the high pseudo-rapidity region.

\begin{figure}[htb]
\begin{center}
\includegraphics[width=.5\textwidth]{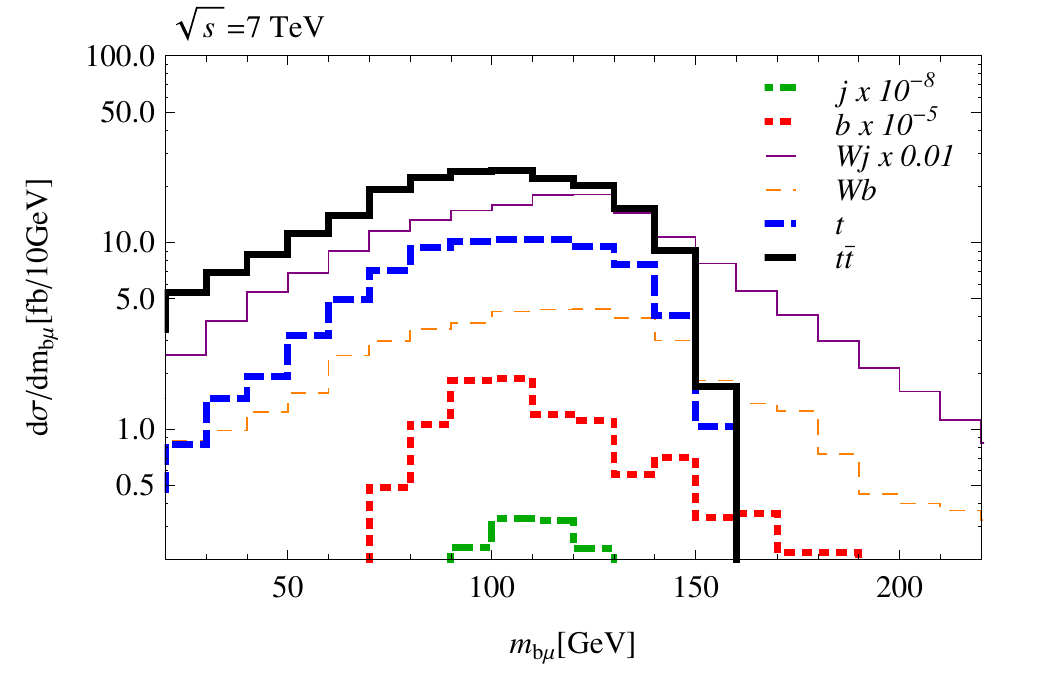}
\caption{The $\ttbar$ signal and background distributions as a function of the invariant mass of the candidate $b$ and muon, $\mbl$, see text for details.  The curves from top to bottom (at $\mbl = 100$ GeV) are for $\ttbar$, $Wj$, single top, $Wb$, $bb$, and $jj$.
}
\label{Fig:SandB}
\end{center}
\end{figure}

\mysection{Signal and backgrounds}
In order to identify
top quarks at LHCb we use their decay $t\rightarrow W b$;
$W\rightarrow\mu\nu$, where both the muon and $b$ have to be
in the acceptance of the detector, defined by the
approximate pseudorapidity range, $2<\eta<5$~\cite{LHCbTDR}.
LHCb provides enhanced
detection of muons with respect to electrons so in the
following for this first study we only consider final state
muons from $W$ decay.
We consider events with large invariant mass and transverse momenta, $p_T$.
Throughout, the muon is required to have $\pt > 20$ GeV and a moderate isolation cut of $\Delta R=0.4$ is imposed, where $\Delta R^2=\Delta \eta^2+\Delta \phi^2$ with $\phi$ being the muon's azimuthal angle, respectively.  Finally, a cut of $\pt>50\,$GeV is imposed on the $b$ jet, which retains most of the signal events.
In Fig.~\ref{Fig:SandB} we show the resulting SM $\ttbar$ signal  (in thick full black line)
as a function of the $b-\mu$ invariant mass, $\mbl\,$.
The signal, as well as the $Wj$, $Wb$, and single top backgrounds (see below), were obtained at leading order,
using MadGraph/MadEvent 4.4.57~\cite{Alwall:2007st}, at the partonic level, and using CTEQ6L1 PDFs~\cite{Pumplin:2002vw}. 
The signal curve has been rescaled by a $K$-factor of 1.7, corresponding to an inclusive $\ttbar$ cross section of
150 pb, within the range obtained at NLO + NNLL order in \cite{Ahrens:2011mw, Kidonakis}, and consistent with recent LHC measurements~\cite{ATLAStt,CMStt}. The LHCb detector response is not included in any of our
simulations.

We divide the backgrounds broadly into two categories: the first one includes a genuine hard muon from a $W$ decay, and the second one involves either a fake or secondary isolated muon from a light flavor jet, $j$ (including charm), or a $b$ quark. In Fig.~\ref{Fig:SandB} we also show the $Wx$ backgrounds belonging to the first category, where $x=b,j$ ($x$ corresponds to the leading jet), and the backgrounds are plotted as functions of $\mxl$  (in thin dashed orange and full purple lines respectively).  A cut of $\pt> 50$\,GeV is imposed on $x$. 
The ATLAS collaboration has recently reported an inclusive $Wj $ cross section times leptonic branching ratio of $0.84$\,nb~\cite{:2010pg}. 
We have rescaled the $Wx$ curves by a $K$-factor of 1.2, which reproduces the central value of the measurement 
under the same set of cuts. A signal to background ratio above one can
be obtained for the $Wj$ background, if a $j\to b$ mistag
rate of $1:100$ can be achieved, while maintaining a large
$b$ jet detection efficiency. The fact that this is in the
ball park of the mistag rates found by ATLAS and
CMS~\cite{CMSpaper, AtlasConfnot} (for a $b$-tagging
efficiency of $50\%$) is encouraging. For charm jets, Êthe
$Wc$ background can be brought to a level at or below the
top signal with a far more modest mistag rate (consistent
with~\cite{CMSpaper, AtlasConfnot}). 
The a priori worrisome $Wb$ irreducible background lies well below the signal.

Single top production, due to its forward nature, is another relevant irreducible background for the $\ttbar$ signal.  
As shown in Fig.~\ref{Fig:SandB} (in thick dashed blue line), within the SM and with the cuts described above, a signal to background ratio of a few is expected.  Our leading order curve for the sum of single top and anti-top production corresponds to an inclusive cross section of 62 pb, consistent with a recent approximate NNLO analysis~\cite{Kidonakis:2011wy}, and a prior NLO analysis~\cite{Schwienhorst:2010je} (we have checked that the $Wt$ contribution~\cite{Tait:1999cf} to the single top signal is negligible). 
Note that single top measurements at ATLAS and CMS, particularly at the high end of their 
pseudorapidity reach, $\eta \sim 2$, will be useful for calibrating single top production in 
the various Monte Carlo tools. A detailed study of the differences between single top and $\ttbar$ events, {\it e.g.}
the presence of a second $b$ jet in the forward direction, may allow a further reduction of the single top background.
It is important to note that the LHCb is sensitive to models in which single top production receives a large forward enhancement 
(see \cite{Craig:2011an} for a recent discussion).


Backgrounds in the second category consist of QCD production of $b\bar b$ as well as light jets, where one jet inside the detector is mistagged as an isolated muon and the other one is identified with a $b$ quark. We have simulated these backgrounds using MadGraph interfaced with Pythia 6.4.14~\cite{Alwall:2008qv} for showering and hadronization.  FastJet~\cite{FastJet} has been employed for jet clustering using the anti-$k_t$~\cite{antikt} algorithm with $R=0.4$.  Cuts of $p_T>50\,$GeV are imposed on the leading $b$ or light jet. For the $jj$ background we assume a $j\to b$ mistag rate of $1:100$, as discussed above.
Fake $j\to \mu$ muons originate from calorimeter punch through and also from early leptonic decays of pions and kaons. The former can be removed with a cut on the maximum energy deposited in the hadronic calorimeters~\cite{thesis}. The muons originating from decay in flight can be efficiently rejected by requiring an isolation cut. We estimate the rejection power 
by requiring that the subleading jet in $p_T$ contains only a single particle (pion or kaon). In addition, we employ an early leptonic decay rate of $10^{-3}$, as obtained with a full detector simulation in~\cite{thesis}. Combining the two
yields a rejection power of $1:10^6$.  For the $b\to \mu$ fake rate we require that one $b$ decays (semi)leptonically and apply a $\Delta R=0.4$ isolation cut on the emitted muon, resulting in a rejection power of $1:10^5$.  
In Fig.~\ref{Fig:SandB}, the raw $jj$ and $bb$ backgrounds (drawn in thick dot-dashed green and dotted red lines respectively) are multiplied by 
$10^{-8}$ and $10^{-5}$, respectively, demonstrating that they are reduced  to levels well below 
the signal using our estimates.
%

As Fig.~\ref{Fig:SandB} shows, after the cuts described above and with a $j\to b$ mistag rate of $1:100$, a signal to background ratio near one
is expected.  However, the largest background, due to $Wj$,
could be well measured given a precise determination of the $j\to b$ mistag rate at LHCb.
Consequently, with enough statistics the $\ttbar$ signal can be extracted. For instance, with the above cuts more than one hundred $\ttbar$ events are expected for one fb$^{-1}$.

\begin{figure}[htb]
\begin{center}
\includegraphics[width=.5\textwidth]{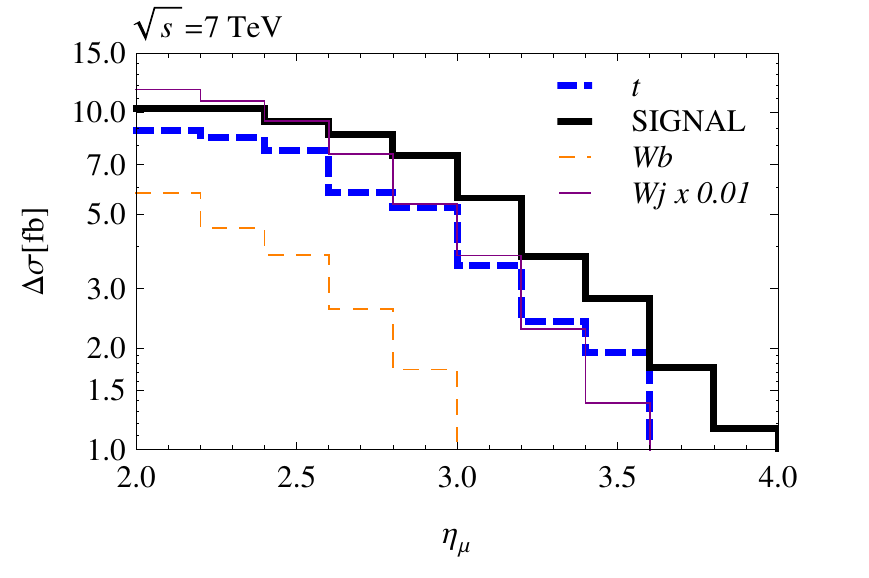}
\includegraphics[width=.5\textwidth]{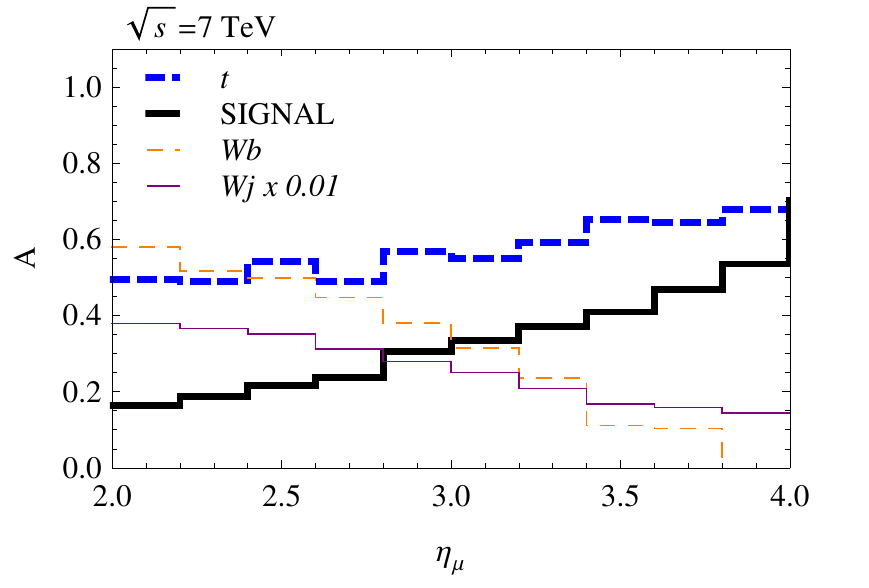}
\caption{The signal and background top anti-top cross section differences (upper pannel) and individual rate asymmetries (lower pannel), as functions of  $\eta_\mu$. See text for details.
}
\label{Fig:eta}
\end{center}
\end{figure}

\mysection{Forward-backward asymmetry}
At the LHC there is a priori no preferred direction of collisions due to the symmetric nature of the initial state. In principle, one can measure a forward backward asymmetry based on the fact that on average the proton's valence quarks carry larger momentum fractions.
Hence, the event boost is correlated with the initial quark direction, leading to a physical axis with respect to which an 
asymmetry could be measured.
Unfortunately, full reconstruction of the event and its boost is not possible at LHCb due to the detector's limited angular coverage.
Instead, we propose a way to indirectly measure the forward-backward asymmetry.
In the absence of an asymmetry, the $\ttbar$ pseudorapidity distribution is symmetric, {\it i.e.}, there is no difference between the top and anti-top distributions as functions of $\eta$. However, a positive forward-backward asymmetry would imply that the top direction is correlated with the $u$ or $d$ parton direction from the hard part of the interaction. Hence it is expected to be more boosted and forward on average, compared to the anti-top. 
Thus, one would expect the forward-backward asymmetry to generate a $\ttbar$ rate asymmetry at given pseudorapidity, 
\beq
\Aet= \left({d\sigma^t/d\eta-d\sigma^{\bar t}/d\eta \over d\sigma^t/d\eta+d\sigma^{\bar t}/d\eta }\right)_{\eta\in 2-5},
\eeq
resulting in a different number of tops vs. anti-tops in the LHCb detector.  
This is demonstrated in Fig.~\ref{Fig:eta}, where the difference between the top and anti-top cross sections (numerator of $\Aet$) as well as the rate asymmetry,  are plotted as functions of the muon pseudorapidity, $\eta_\mu$ (alternatively, one could also study the dependence on the $b$ pseudorapidity). 
For illustration,  the NP signal (drawn in thick full black line) is due to 
$t$-channel $Z^\prime$ exchange, see Jung {\it et al.} in \cite{ttNPtchannel}, with parameters chosen to yield a sizable forward-backward asymmetry in the forward region ($\ad_{\dy>1}  = 0.43$ at leading order in QCD).  The SM leading order contribution is symmetric, consistent with no rate asymmetry. 

The $Wj$, $Wb$, and single top backgrounds also yield a rate asymmetry. Their impact is included in Fig.~\ref{Fig:eta} (in thin full purple, dashed orange and thick dashed blue lines respectively), where the actual rate differences and the individual asymmetries are shown in the upper and lower panel, respectively.    
The largest background to the top anti-top cross section difference 
is due to $Wj$ (again we have assumed a $j\to b$ mistag rate of $1:100$).
However, the underlying $Wj$ cross section asymmetry should be be well measured 
by LHCb, due to the large statistics that will be available
in $Wj$.  Thus, precise knowledge of the $j\to b$ mistag rate would accurately determine this background
for $\Aet$.  Sizable contributions to $\Aet$ are also expected to arise from single top production, see Fig.~\ref{Fig:eta}. 
Our single top simulation corresponds to inclusive cross sections of 41 pb ($t$) and 21 pb ($\bar t$), consistent with~\cite{Kidonakis:2011wy,Schwienhorst:2010je}.
Note that precise ATLAS and CMS measurements of the $Wj$ and single top cross section asymmetries
at lower pseudorapidities will again be useful for calibrating the relevant Monte Carlo tools.

%


We emphasize that our analysis does not aim to replace a state of the art experimental effort, including optimization of cuts and detector effects. We merely wish to point out that such an analysis may be feasible and worthwhile, especially if the NP leads to anomalous top kinematics in the forward direction. 
Finally,  we note that the $\pt$ and pseudorapidty distributions of the muon~\cite{ptlepton}, which is known to be a perfect top-spin analyzer, may provide LHCb with sensitivity to differences between the polarization of the top produced in the SM and in its extensions. Moreover, a measurement of the top polarization could shed light~\cite{Jung:2010yn}  on the origin of a large top charge asymmetry in the forward region. 

\mysection{Acknowledgments}
We thank Ohad Silbert for useful discussions.
A.L.K. is supported by DOE grant FG02-84-ER40153.
J.F.K. is supported in part by the Slovenian Research Agency. 
 G.P. is the Shlomo and Michla Tomarin career development chair and supported by the Israel Science Foundation (grant \#1087/09), EU-FP7 Marie Curie, IRG fellowship, Minerva and G.I.F., the German-Israeli
Foundations, and the Peter \& Patricia Gruber Award. 
S. Stone thanks the U. S. National Science Foundation for support.

\end{document}